# Video Analysis for Body-worn Cameras in Law Enforcement

Jason J. Corso, University of Michigan
Alexandre Alahi, Stanford University
Kristen Grauman, University of Texas at Austin
Gregory D. Hager, Johns Hopkins University, CCC
Louis-Philippe Morency, Carnegie Mellon University
Harpreet Sawhney, SRI
Yaser Sheikh, Carnegie Mellon University

June 2015

The social conventions and expectations around the appropriate use of imaging and video has been transformed by the availability of video cameras in our pockets. The impact on law enforcement can easily be seen by watching the nightly news; more and more arrests, interventions, or even routine stops are being caught on cell phones or surveillance video, with both positive and negative consequences.

This proliferation of the use of video has led law enforcement to look at the potential benefits of incorporating video capture systematically in their day to day operations. At the same time, recognition of the inevitability of widespread use of video for police operations has caused a rush to deploy all types of cameras, including body worn cameras. However, the vast majority of police agencies have limited experience in utilizing video to its full advantage, and thus do not have the capability to fully realize the value of expanding their video capabilities.

The use of body-worn cameras has the potential to provide many broad advantages to law enforcement. Some examples are:

- **Transparency:** the willingness to video day-to-day activities and release the video to the public has the potential to increase public trust and confidence in the police.
- **Officer protection**: body-worn cameras can protect officers from false allegations and influence the behavior, in a positive way, for both the officer and those being recorded.
- **Investigative**: capturing spontaneous events, crime scenes, etc. will aid in the investigation of crimes and the prosecution of these cases. The cameras supplement the officer's recall and document events.
- **Training**: the recorded real-life situations will aid in educating both green and experienced officers.

President Obama recently proposed a *Community Policing Initiative* that would provide a 50% match to states/localities who purchase body-worn cameras and requisite storage. There are estimated to be about 850,000 law enforcement personnel in the United States. Many agencies have rushed to deploy body-worn cameras and this will continue for the foreseeable future. A great majority of these agencies cobbled together a solution using local personnel and frequently underestimated both the complexity and cost of operating a body camera system.  As is detailed below, numerous immediate technology needs of the law enforcement communities are beyond the available current technology.

In this white paper, we highlight some of the technology needs and challenges of body-worn cameras, and we relate these needs to the relevant state of the art in computer vision and multimedia research.  We conclude with a set of recommendations.

**Technology Drivers**

Body-worn cameras present many advantages to law enforcement, as noted above. However, in order to effectively address these needs, they will need to be supported by practical tools that allow law enforcement agencies, already often resource limited, to make effective use of video. Below, we describe specific technology drivers that will need to be addressed in order to realize these advantages.

> **Redaction of sensitive materials-** Redaction is cited as one of the most urgent needs for the police departments adopting bodycams. Traditional redaction, e.g., as used by Google Street View to blur faces and license plates, is only a first step. Sensitive identifying information can take many subtler forms in both visual and audio streams, including, for example, a logo on a shirt, or contextual cues that indicate specific locations. Hence, both basic person identification methods as well as methods able to detect more nuanced identifying cues are needed.
>
> **Freedom of Information Act (FOIA) servicing-** Redaction is only part of the concerns or technology needs for servicing Freedom of Information Act queries. Certain queries can request video segments that satisfy various criteria, such as time of day, number of officers present, etc.  Furthermore, different jurisdictions have different FOIA laws about what types of requests are permissible. These present high-cost challenges to local police departments. By the account of the Dallas Police Asst. Chief Lawrence, the Dallas PD will need to double the number of personnel it has to handle FOIA requests, for example.
>
> **Forensic search and triaging-** Related to FOIA servicing as well as investigative needs, the ability to index, search, and triage large repositories of body camera video footage will be a critical forensic capability.  These forensic capabilities will require the ability to analyze data at varying levels of specificity, e.g., from geospatially and temporally localized footage to general queries about a certain type of activity.
>
> **Training systems-** The data collected through wearable sensors provide realistic videos and other types of data for training law enforcement personnel with genuine context and scenarios.  These videos will need to be curated and matched to certain training needs and scenarios.

**Early warning systems-** There is a need to better capture data on how any given officer is interacting with the public, and detect early warning signs, such as repeated premature use of force. Currently, the officer largely self-reports this data (e.g., about use of force, whether there was a foot chase, etc.). In the future, the bodycams could enable more automated collection of such data. This can help detect triggering behaviors before they escalate to larger problems.

Finally, although not a direct technology driver, the need for **standards** to be established and practiced is paramount to the realization of these technology drivers. Standards crosscut all aspects of body-worn cameras in law enforcement, including usage practices, technology standards, and both research and commercial growth in this space.

**Technology Challenges and Enablers**

The computer vision and multimedia communities have made large strides in recent years leading to such capabilities as face detection in consumer cameras, automated audio/video indexing and retrieval, and automated speech recognition systems. Although promising, these advances have largely come in discriminable content domains such as various Olympic sports or recognizing actor pose in certain professionally-shot movies. Furthermore, the level of performance for typical large datasets does not meet certain body camera technology needs, such as FOIA servicing in which even a single missed face or license plate can be costly.

More recently, the computer vision community has begun to look at body-worn camera footage, which we call "egocentric vision" or "first-person vision". Some work addresses recognition tasks from the first person point of view, particularly for activity recognition. It is typically assumed that activities are defined by the objects and people with which the camera wearer is interacting, e.g., detecting the activity of doing laundry entails detecting the washing machine and the clothes. In traditional third-person vision there is ample research on understanding human activities and interactions, e.g., in surveillance style video, and there are early steps towards addressing similar tasks from the first person perspective.

Another problem that researchers in our community have begun to focus on is summarizing long first-person videos. A person with limited time needs to be able to quickly understand the content of a long video without having to watch the full sequence. This is especially true in the case of egocentric video, which is likely to contain long periods of uninteresting footage while driving, walking, etc. There is thus a critical need for video summarization methods to deliver the information content of long videos in a short period of time. While the need for egocentric video summarization methods is clear, the development of such methods within the computer vision community has been hampered by the lack of a standard, efficient way to evaluate video summaries. This lack can be attributed to an ill-posed highly subjective goal such as "informativeness". The utility of a summary is likely to be somewhat linked to the exact application domain where it is being used; meaning a good summary for law enforcement applications has (potentially) distinct properties from a good summary for the consumer life-logger domain. We have hence seen continued interest in automatic indexing of semantic content within these videos, such as detection of actions and events for browsing and retrieval tasks.

In contrast to the predominant data used in the research community, video from body-worn cameras presents new challenges to the computer vision research community. The video is shaky due to rapid movements of the officer. The video is sometimes not focused on the scene at large but rather at nearby objects, even the ground, in certain situations.

Perhaps more importantly, audio and video carry important and complementary content. However, the challenges for audio interpretation are similarly large. Audio is often washed out by loud environmental noise. The audio often has concurrent multiple speakers. The joint video and audio streams are capturing rapid movements of both officers and non-officers. There are high degrees of occlusion and articulation in these interactions.

Orthogonally, the practical challenge of sufficient battery life and recording technology is critical to minimizing usage barriers. Effective use of the body-worn cameras require operation through a full shift without recharging of batteries or off-loading of video. Current estimates for battery life on body-worn cameras are 8-12 hours of normal use (not continuous capture and storage). Estimates for initial police department usage suggest each officer wearing a body-worn camera will produce about 3 hours of video per day.

Finally, two crosscutting challenges relate to the acquisition and use of data by the research community. Current video acquisition and storage systems are evolving rapidly, but are doing so within the context of closed and proprietary platforms. Access is controlled through proprietary interfaces, and the vendor defines data formats and metadata formats. As a result, it is both difficult to pull and share data from these systems, and equally challenging to share data across systems. Consistent standards and more open platforms will be key to driving innovation.

The second crosscutting challenge is the need to bootstrap development using curated and labeled video. Initial conversations suggest that there are already weak labels for much of the video "of consequence" that is captured every day. However, providing more early guidance on the structure of this labeling process may provide a much more powerful data platform to build on for the future.

The challenge in bridging from the advances recently made to meet the technology needs and challenges of body-worn cameras is significant. Given the state of the art in computer vision and multimedia, and the technology challenges presented by body-worn cameras in law enforcement, we expect the following technology enablers indexed over 2-, 5-, and 10-year timelines.

**2-Year Timeline**

The first group of technology enablers will require the least amount of refinement from current methods. We expect that within two years, the underlying technology for these capabilities could be finished. However, transitioning them to functioning tools for end-user adoption will require investment (see Recommendation section for more discussion on this point).

- Establish data standards for video and audio as well as associated metadata; encourage open standards and platforms to promote minimal technical barriers to data export, access, and exchange.
- Develop a canonical set of actions/behaviors of relevance and importance to law enforcement. These common scenarios would be released across the nation to facilitate translation across jurisdictions.
- Automatic detection and redaction of certain classes of entities such as faces, logos, license plates, etc., and automatic detection and indexing of certain specific types of events such as fires, crowds, explosions. These automatic detections would be mapped to the canonical semantics established above.
- Query-based search through video sets for certain classes of entities or specific entities. These would involve search by entity-type or search by example.
- Automatic acoustic privacy filters to mask various environmental sounds.
- User interfaces for (1) video annotation as they are captured and post-processed, (2) redaction and (3) search. Voice annotation is a likely target, especially for annotation during capture. The user experience with these systems is critical for their adoption and use. Novel human-machine interaction methods, including active learning, can be employed.
- Reality-based training in familiar environments for which models of entities, actions and locales are pre-defined. The modules could, for example, leverage actual footage retrieved by the semantics and search mechanisms defined above.
- Creation of representative data corpora to facilitate common benchmarking and new methodological development of key problems like automatic redaction, search and indexing.
- Automated methods based on audio or simple (lightweight computationally) visual signals to decide when to turn the camera on or when it may be safely turned off.
- Video summarization to improve an officer's memory of a certain event he/she experienced. Examples include short keyframe or video skim summaries. Simple acceleration of the video, if combined with video stabilization (e.g., using the commercially available Microsoft Hyperlapse approach) is another possibility to ease watching of long first-person video.

**5-Year Timeline**

The second group of technology enablers will require further foundational and applied research and technology development. As the list below indicates, these build on the earlier 2-year timeline points.

- Improvement of the automatic detection and redaction of certain classes of entities to include performance levels requiring minimal or no manual verification; to include more nuanced fine-grained and contextual aspects of detection and redaction including but not limited to store-fronts, street addresses, specific vehicles, specific ages of people (e.g., juvenile).

- Extension of the automatic detection and redaction of certain classes of events toward the ability to detect novel complex events as simple compositions of pre-specified ones.
- Forensic indexing and search enhancement by fusing multiple camera accounts of common events.
- Limited real-time detection and display of certain entities toward situational awareness.
- Inter-connected front-end (officer in the field) and back-end (HQ) to facilitate dynamic interaction among systems. Connecting multiple officers in the field is also possible.
- Reality-based training with both virtual and augmented reality, using example body-worn video footage that has been stored, processed and indexed using the earlier developments.
- Video and audio encoding and compression that is customized to the particular forensic need and use of body-worn cameras rather than the use of off-the-shell encoding and compression algorithms that are currently used.
- Personalization of body-worn cameras to officer-specific roles and behaviors. There is rich potential in learning from the long history of captured footage for a particular officer; with minimal human input, the interface and processing backend can be tailored to his or her needs, preferences and practices.

**10-Year Timeline**

The third group of enablers are far-reaching ideas with the potential for significant impact. In turn, they require sizeable initial basic research investment.

- Real-time redaction of footage according to established policies. Adaptable redaction of footage as a function of the context of use; e.g., blurring of storefronts to protect location information when needed.
- A level of full-scale automatic situation awareness is possible. Officers in the field will be in peer-to-peer communication with each also connected to the HQ. Dynamic and tuned situational awareness will be derived in a mix of front-end/real-time technologies and back-end technologies. Situational awareness will be communicated to relevant officers.
- Large-scale indexing and searching through the body-worn camera footage with queries specified in a combination of natural language and visual examples.
- Body-worn camera summarization systems able to compress 3-24 hours (a day to a week of officer activity) of video into short human interpretable visual "stories", that can be textually read/indexed, visually-textually read like comic-books, or visually-audially watched like short movies.
- Data mining of large archives of multiple officers' data to detect trends in behavior of both officers and members of the public, both for training and forensic purposes.
- Real-time monitoring of the officer's state in terms of emotional health, alertness, etc.

## Recommendations

We make the following recommendation to policy makers, to police departments planning to deploy body-worn cameras on officers, and to system manufacturers planning to engage the need.

**Policy Recommendations**

    **Usage Protocol-** We recommend that policy makers develop and provide police departments and system manufacturers with recommended best practices for operating body worn sensors. The standardization of a usage protocol will structure officers' reactions and will simplify automated indexing and search analysis. This protocol should include recommendations on when to turn devices on, how much history to buffer, guidelines for narration and final debriefing, and a clear explanation of how the captured data will be used.

    **Public Education-** In conjunction with educating police officers on the proper use of sensors, we recommend developing a plan for educating the public and journalists on how to access and correctly draw conclusions from the data. For instance, it is not possible to guarantee the camera viewpoint is that of the officer, with body-worn cameras. Therefore, there is no good reason to limit the capture viewpoint capabilities. We, in fact, recommend storing a wide viewpoint and high resolution. Educate media and public as to this fact.

**Technology Recommendations**

    **Multimodal Sensing-** Federal and state leadership should provide minimal sensor guidelines for audio, visual, and metadata sensors. We recommend a stereo pair of wide field-of-view, high-resolution cameras, microphones with sufficient dynamic range to capture human speech, an in-built inertial measurement unit, and GPS. All these sensors should be precisely time-stamped to a GPS-locked clock.

    **Media Central-** Two immediate issues that are expected to arise are data storage and differentiated access to analytics technology. We recommend the development of a central media facility that police departments across the country may avail to store and analyze their data. The system should be cloud-based and provide state-of-the-art tools in indexing and searching video. Future functionalities that will be enabled by research should be incorporated into the media central facility. This setup will assist in disseminating the latest advances in analytics across all police stations. Privacy concerns will need to be adequately addressed.

    **Indexing-** Data storage and indexing needs are evident. The data should be indexed against the state of the art visual and audio indexing technologies. The data should be stored indefinitely. This long-term storage facilitates later processing and reprocessing of the data against new analytics technologies. Both the visual signals

and the audio signals are valuable sources of information for generating a complete understanding of a given situation.

**Research Recommendations**

**Standards-** Cultivate a competitive ecosystem for startups. Ensure data is consistent and available across jurisdictions. A primary risk factor for widespread **effective** adoption of body-worn cameras and the new capabilities they enable is the creation of a set of non-interoperating closed systems that do not allow for innovation, research, or development on this data.

**Research funding-** The Community Policing Initiative funding provides financial support for the acquisition of body-worn cameras and requisite storage. However, it does not account for the many unsolved research questions presented herein, nor does it account for the expected high cost of new personnel to manage and make use of the data (e.g., to service FOIA requests). Without funding for new research programs and technology transition (below), our primary concern is that these highly informative information sources will essentially be wasted. Research funding, both basic and applied, is hence needed to facilitate the best use of these novel highly informative information sources.

**Technology Transition-** A primary obstacle to realizing the possibilities of body-cameras is the transition from university and industrial lab to practice. Prototype systems that result from research programs in the lab are unsuitable for direct transition to the end-user. Further transition funding to get these developments into the hands of the end user is necessary. Traditional transition funding from various federal agencies seems insufficient to support such a transition. We hence underscore the need for open standards and interoperability to cultivate an ecosystem for startups and other private investments in the sector.

**Datasets and Benchmarks-** Much recent progress in the computer vision and multimedia communities has been driven by the availability of large, expert-annotated data sets. Such data sets create a level playing field for researchers and developers in the respective problem-spaces and force methodologies to scale to the level of difficulty of the data set. Research in body-worn cameras is no different. Desiderata of such data sets include: annotation (e.g., of to-be-redacted entities), size (e.g., training and testing on the wide-range of scenarios), the same general characteristics of the end-game scenarios, large, real-world video and audio (i.e., not acted whenever possible). The computer vision community has significant expertise in creating and curating such data sets and should be intimately involved.

**Continued Involvement Among Video Processing Research Community-** Some mechanism for establishing immediate and continued involvement of the academic and industrial research communities is needed. In the near term, a larger workshop involving the law enforcement and the research community is needed. Longer-term possibilities are a cross-disciplinary panel, or, relating to the standards point above, a standards committee.

**Conclusion**

The public trust in the safety and security provided by our collective law-enforcement agencies is of paramount importance. We have outlined a set of technology drivers, challenges, and enablers surrounding the use of body-worn cameras by law enforcement, and we have made a set of recommendations that we believe are critical to the effective use of body-worn cameras. The outline and recommendations resulted from a panel of computer vision experts and law enforcement personnel and subsequent discussions.

**Acknowledgements**

This white paper grew out of a meeting held at the IEEE Conference on Computer Vision and Pattern Recognition and sponsored by the CCC. Attendees at the meeting were Alexandre Alahi (Stanford), Ann Drobnis (CCC), Greg Hager (JHU), Harpreet Sawhney (SRI), Jason Corso (U of Michigan), Jill Crisman (IARPA), Joanna Weiss (Arnold Foundation), John Garofolo (NIST), Kristen Grauman (U of Texas Austin), Louis-Philipe Morency (CMU), Mubarak Shah (U Central Florida), Randy Bryant (OSTP), Assistant Chief Thomas Lawrence (Dallas Police Department), and Yaser Sheikh (CMU). Dr. Liz Shriberg (SRI), Dr. Gerald Friedland (ICSI at UC Berkeley), and Prof. Raymond Surette (U Central Florida) also contributed to the white paper via e-mails, discussions and references.

*For citation use: Corso J. C., Alahi A., Grauman K., Hager G. D., Morency L., Sawhney H., & Sheikh Y. (2015). Video Analysis for Body-worn Cameras in Law Enforcement: A white paper prepared for the Computing Community Consortium committee of the Computing Research Association. http://cra.org/ccc/resources/ccc-ledwhitepapers/*

This material is based upon work supported by the National Science Foundation under Grant No. (1136993). Any opinions, findings, and conclusions or recommendations expressed in this material are those of the authors and do not necessarily reflect the views of the National Science Foundation.